\documentclass{IEEEtran4PSCC}

\makeatletter
\def\input@path{{./tables/}}
\makeatother
\usepackage{booktabs}

\newcommand{\ff}[1]{$\mathcal{F}_{#1}$}
\newcommand{\xxx}[1]{$(#1_{1},\,#1_{2},\,#1_{3})$}

\usepackage{amssymb}
\usepackage{lipsum}

\ifCLASSINFOpdf
   \usepackage[pdftex]{graphicx}
\graphicspath{{./figures/}}
\else
   \usepackage[dvips]{graphicx}
\fi
\usepackage[cmex10]{amsmath}
\usepackage[caption=false,font=footnotesize]{subfig}

\makeatletter
\let\old@ps@headings\ps@headings
\let\old@ps@IEEEtitlepagestyle\ps@IEEEtitlepagestyle
\def\psccfooter#1{%
    \def\ps@headings{%
        \old@ps@headings%
        \def\@oddfoot{\strut\hfill#1\hfill\strut}%
        \def\@evenfoot{\strut\hfill#1\hfill\strut}%
    }%
    \def\ps@IEEEtitlepagestyle{%
        \old@ps@IEEEtitlepagestyle%
        \def\@oddfoot{\strut\hfill#1\hfill\strut}%
        \def\@evenfoot{\strut\hfill#1\hfill\strut}%
    }%
    \ps@headings%
}
\makeatother

\psccfooter{%
        \parbox{\textwidth}{\hrulefill \\ \small{22nd Power Systems Computation Conference} \hfill \begin{minipage}{0.2\textwidth}\centering \vspace*{4pt} \includegraphics[scale=0.06]{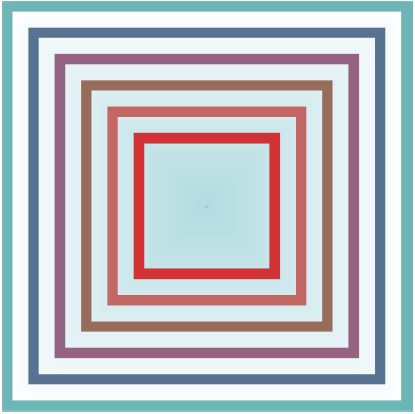}\\\small{PSCC 2022} \end{minipage} \hfill \small{Porto, Portugal --- June 27 -- July 1, 2022}}%
}

\begin{document}

\title{Design and Operation of Hybrid Multi-Terminal\\Soft Open Points using Feeder Selector Switches\\for Flexible Distribution System Interconnection}


 \author{\IEEEauthorblockN{Matthew Deakin\IEEEauthorrefmark{1},
 Phil C. Taylor\IEEEauthorrefmark{2},
 Janusz Bialek\IEEEauthorrefmark{1}, and
 Wenlong Ming\IEEEauthorrefmark{3}}
 \IEEEauthorblockA{\IEEEauthorrefmark{1}School of Engineering\\
 Newcastle University, 
 Newcastle-upon Tyne, UK}
 \IEEEauthorblockA{\IEEEauthorrefmark{2}Faculty of Engineering\\
 University of Bristol,
 Bristol, UK}
 \IEEEauthorblockA{\IEEEauthorrefmark{3}School of Engineering\\
 Cardiff University,
 Cardiff, UK}
 }

\maketitle

\begin{abstract}
Distribution systems will require new cost-effective solutions to provide network capacity and increased flexibility to accommodate Low Carbon Technologies. To address this need, we propose the Hybrid Multi-Terminal Soft Open Point (Hybrid MT-SOP) to efficiently provide distribution system interconnection capacity. Each leg of the Hybrid MT-SOP has an AC/DC converter connected in series with a bank of AC switches (Feeder Selector Switches) to allow the converter to connect to any of the feeders at a node. Asymmetric converter sizing is shown to increase feasible power transfers by up to 50\% in the three-terminal case, whilst a conic mixed-integer program is formulated to optimally select the device configuration and power transfers. A case study shows the Hybrid MT-SOP increasing utilization of the converters by more than one third, with a 13\% increase in system loss reduction as compared to an equally-sized MT-SOP.
\end{abstract}


\thanksto{\noindent Submitted to the 22nd Power Systems Computation Conference (PSCC 2022). This work is supported by Supergen Energy Networks hub (EP/S00078X/2) and Multi-energy Control of Cyber-Physical Urban Energy Systems (MC2) project (EP/T021969/1). Corresponding email: \texttt{matthew.deakin@newcastle.ac.uk}}

\section{Introduction}

To reach Net-Zero, it is imperative that carbon intensive components of energy systems are replaced with Low Carbon Technologies (LCTs) such as solar PV, wind generators, heat pumps and electric vehicles. These LCTs are often connected directly to distribution networks, resulting in a need for increased network capacity, as well as presenting an opportunity to re-imagine network operations to reduce whole system costs. Network capacity has traditionally been provided with reinforcement, whilst lifetime costs are managed at the planning stage using the fit-and-forget approach (e.g., through the installation of low-loss assets).

A promising alternative to these conventional reinforcement approaches is via the provision of flexible, controllable interconnection capacity between radially operated feeders. Normally Open Points (NOPs), located at end points of distribution feeders, can be replaced with Soft Open Points (SOPs) \cite{bloemink2010increasing,fuad2020soft} to provide this capacity. Where two feeders meet at a NOP, back-to-back voltage source converters can be installed to form a SOP, allowing power to be transferred whilst maintaining radiality. When three or more feeders meet at a NOP, a Multi-Terminal SOP (MT-SOP) replaces the NOP. SOPs and MT-SOPs can provide network capacity and opportunities for reduced operating costs \cite{fuad2020soft}. Unfortunately, power converters continue to be expensive on a power-rated basis \cite{huber2017applicability}, and so finding approaches to maximize the utilization of these converters will therefore be paramount for the potential of SOPs and MT-SOPs to be realized.

In this context, prior works have focused on the operation of SOPs and MT-SOPs given fixed connections to distribution network feeders \cite{fuad2020soft}. For example, in \cite{sun2020optimized}, the authors propose an approach for optimal dispatch of a multi-terminal SOP in a robust manner whilst considering distribution system unbalance. In \cite{lyu2021general}, a state-space model of a general MT-SOP is determined for transient analysis, with a three-terminal case study illustrating the approach. The authors of \cite{ji2017enhanced} demonstrate that an MT-SOP can be used for feeder load balancing, thereby avoiding thermal overload of branches in a network. Where reconfiguration of networks is considered, it is in the context of traditional tie-line reconfiguration (e.g., \cite{bai2018distributed}).

\begin{figure*}\centering
\subfloat[Equally sized MT-SOP]{\includegraphics[width=0.32\textwidth]{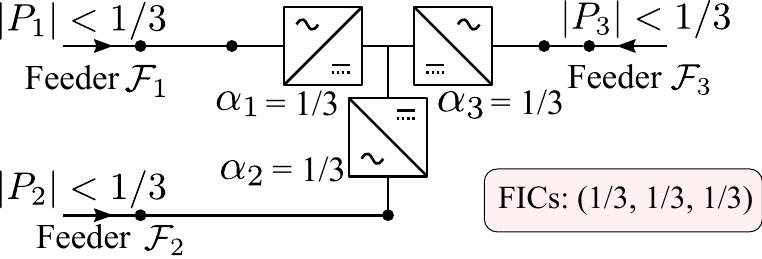}\label{f:intro_A}}~
\subfloat[Asymmetrically sized MT-SOP, configuration~1]{\includegraphics[width=0.32\textwidth]{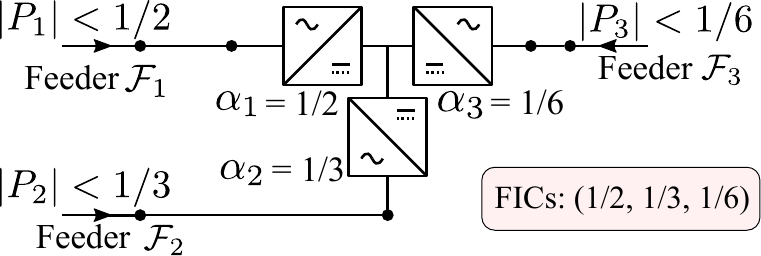}\label{f:intro_B}}~
\subfloat[Asymmetrically sized MT-SOP, configuration~2]{\includegraphics[width=0.32\textwidth]{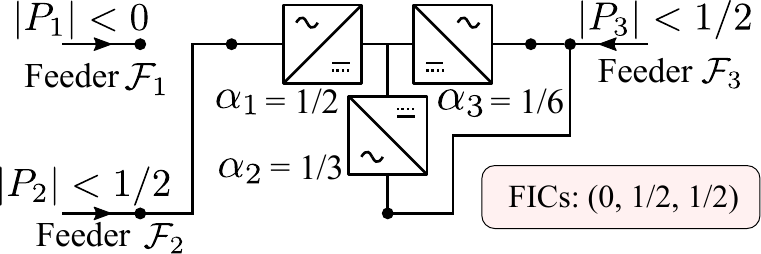}\label{f:intro_C}}
\caption{If the converters sizes \xxx{\alpha} of a 3-terminal MT-SOP are equal, then the power transferred to the DC bus from any feeder \ff{j} can be no greater than 1/3 of the total power rating, FIC capacity is 1/3~pu (a). By contrast, asymmetric sizing allows for up to 1/2~pu to be transferred between feeders (b). Depending on the converter configuration, the FICs for an asymmetric MT-SOP can change (c).}
\label{f:intro_figure}
\end{figure*}

An alternative approach to maximize converter utilization is via the reconfiguration of the power electronics themselves as part of Hybrid AC/DC systems, given the low cost of legacy AC technologies in comparison to their DC counterparts. In \cite{lou2020new}, the authors present a phase-changing SOP, whereby individual legs of a three-phase SOP are permuted (e.g., phase A connected to phase B) to reduce unbalance and improve network performance. The advantages of reconfigurable, parallel AC and DC lines are shown in \cite{shekhar2019boundaries}. Power converters are connected in series with an `MVAC switchyard' in \cite{majumder2021distribution}, presenting a SOP-like Hybrid AC/DC device in which each converter can be connected to one of multiple feeders to provide headroom. These works show that system performance can be improved by reconfiguration of power converters with respect to the distribution system. The design (sizing) of a Hybrid MT-SOP and the impacts on device flexibility and subsequent operational capabilities have not been considered, however. Given the well-known need for system flexibility, this is a significant gap.

In this paper, we consider the design and operation of a three-terminal, Hybrid MT-SOP, with the goal of increasing the power that can be transferred between feeders and subsequently reduce operating costs. It is demonstrated that a bank of low-cost electromechanical switches (the multiport `Feeder Selector Switch'), combined with asymmetrically sized AC/DC converters, can increase the power that can be transferred between feeders by up to 50\% for a three-terminal Hybrid MT-SOP. A conic mixed-integer program is developed to demonstrate a reduction in ongoing operational costs.

This paper is structured as follows. In Section~\ref{s:mop_design}, the principle of operation of the Hybrid MT-SOP is presented, based on the Feeder Selector Switch and asymmetric converter sizing, illustrating how different Hybrid MT-SOP designs can increase the flexible power transfers that can occur between feeders. A conic mixed-integer program is proposed in Section~\ref{s:mop_operation} for determining the optimal operation of the Hybrid MT-SOP whilst taking into consideration the additional device complexity. A case study illustrates the benefits and operation of the device in Section~\ref{s:casestudy}, before salient conclusions are drawn in Section~\ref{s:conclusions}.

\section{Hybrid MT-SOP Operating Principle for Flexibility}\label{s:mop_design}

The purpose of the Hybrid MT-SOP is to increase the utilization of power converters in the provision of flexibility via feeder interconnection, ultimately improving the cost-effectiveness of the device. This interconnection capacity could be used to address constraints (i.e., to address voltage or thermal constraints) or improve operating performance (e.g., reducing losses or unbalance). It is based on the assumption that AC switches are relatively inexpensive in comparison to power electronics capacity. 

In this section, we introduce the asymmetric sizing approach and the Feeder Selector Switch, to illustrate how these innovations can increase the flexibility afforded by a given Hybrid MT-SOP design. For clarity, we consider the case of a three-converter, three-feeder distribution system in this paper. Future works could consider any number of feeders and converters.

Throughout the paper, we make use of Matlab-style notation (i.e., $A[n]$ represents the $n$th element of $A$, with indices starting at 1). In this section, powers are in per-unit, using the total power converter capacity $P^{+}$ as the power base.

\subsection{Principle of Operation}
To illustrate the advantages of asymmetric sizing and reconfiguration of power electronics, in Fig.~\ref{f:intro_figure} we calculate the Feeder Interconnection Capacity (FIC) considering the connection of 1~pu of power electronics between three feeders. For the power of the $j$th feeder, this FIC is defined as the maximum power that can be transferred from a given feeder $P_{j}$, i.e.,
\begin{equation}\label{e:fic}
\mathrm{FIC}_{j} = \max_{(P_{1},\,P_{2},\,P_{3})} |P_{j}| \,.
\end{equation}
For an equally-sized 3-terminal MT-SOP, the FIC for all feeders is 1/3~pu (Fig.~\ref{f:intro_A}). However, by changing the sizes of the individual converters, the MT-SOP FICs change (Fig.~\ref{f:intro_B}), allowing for increased transfer from a given feeder. This does, however, have the effect of reducing the FIC of the other feeders. Should power need to be transferred between any two feeders, reconfiguration of the outputs of the power converters allows for an increased FIC between those feeders (Fig.~\ref{f:intro_C}).

\begin{figure}\centering
\includegraphics[width=0.39\textwidth]{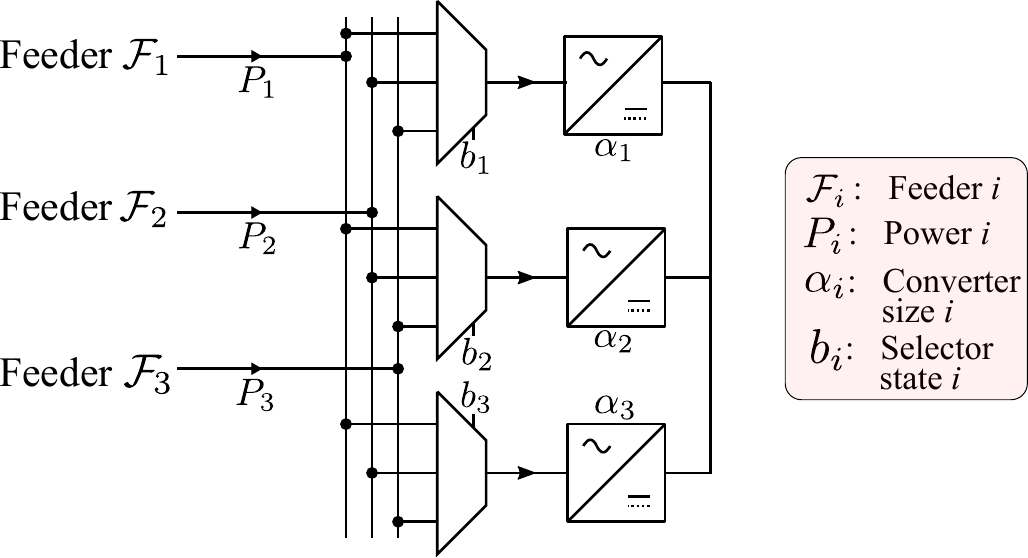}
\caption{Single line diagram of the proposed Hybrid MT-SOP, using asymmetric sizing of converters and Feeder Selector Switches, for the three-terminal case. The Feeder Selector Switch is represented by the multiplexer symbol.}
\label{f:mop_schematic}
\end{figure}

To achieve this reconfiguration, in this paper we propose the Hybrid MT-SOP, outlined in Fig.~\ref{f:mop_schematic}. Each of the power converters is connected to a multi-terminal Feeder Selector Switch, which connects at most one of the feeders to the output. This Feeder Selector Switch is analogous to a multiplexer in telecommunications, static transfer switches used in load balancing in LV distribution \cite{shahnia2014voltage}, or switchyards in multi-terminal HVDC systems \cite{majumder2016alternative}. Compared to the MVAC switchyard described in \cite{majumder2021distribution}, the proposed design explicitly considers the combination of the selector switch with multiple, asymmetrically sized power converters.

\subsection{Hybrid MT-SOP Capability Charts}
The asymmetric sizing and Feeder Selector Switch aspects of the proposed Hybrid MT-SOP have different effects on the locus of feasible power transfers (i.e., the device capability chart), and so these are now introduced in turn.

\subsubsection{Capability Chart of Asymmetric MT-SOP}

For an MT-SOP with asymmetric converter sizes \xxx{\alpha} but without Feeder Selector Switches, then the capability chart for a unity power factor device can be described by the expressions
\begin{align}
|P_{1}|\leq \alpha_{1},\,\quad |P_{2}| \leq &\alpha_{2},\, \quad |P_{3}| \leq \alpha_{3}\,, \label{e:individuals}\\
P_{1} + P_{2} + P_{3} &= 0\,,\label{e:power}
\end{align}
with \eqref{e:individuals} representing the ratings of each converter, and \eqref{e:power} representing the power balance on the DC bus. It can be seen that the FIC capacity \eqref{e:fic} can be calculated for the $j$th leg of the device as
\begin{equation}\label{e:fic_calc}
\mathrm{FIC}_{j} = \min \{\alpha_{j}, \sum _{i \in \mathcal{K}} \alpha_{i} \}, \, \quad \mathcal{K}=\{1,\,2,\,3\}\setminus \{j \}\,.
\end{equation}
For example, for the first leg, $\mathrm{FIC}_{1}$ is $\min \{\alpha_{1}, \alpha_{2}+\alpha_{3}\}$. From \eqref{e:fic_calc}, it can be seen that the equally-sized case does not lead to increased FIC; additionally, it can be seen that the maximum FIC of any design is 1/2~pu.

\begin{figure}\centering
\subfloat[Homogenous sizing]{\includegraphics[width=0.16\textwidth]{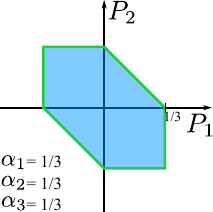}\label{f:cases_1a}}
~
\subfloat[Asymmetric sizing]{\includegraphics[width=0.16\textwidth]{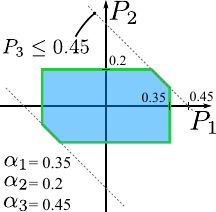}\label{f:cases_1b}}
~
\subfloat[Asymmetric sizing \& Feeder Selector Switch]{\includegraphics[width=0.16\textwidth]{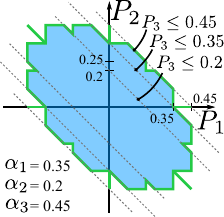}\label{f:cases_1c}}
\caption{A capability chart illustrates the additional flexibility that the Hybrid MT-SOP's asymmetric sizing and Feeder Selector Switch bring. Asymmetric converter sizing increases the FIC capacity to 0.45, whilst the Feeder Selector Switches expands the capability chart.}
\label{f:cases_1}
\end{figure}

To visualize the locus of possible power transfers, we project \eqref{e:individuals}, \eqref{e:power} onto the $(P_{1},\,P_{2})$ plane, to form the capability chart
\begin{equation}\label{e:cap_chart0}
|P_{1}|\leq \alpha_{1},\quad |P_{2}| \leq \alpha_{2},\quad |P_{1}+P_{2}| \leq \alpha_{3}\,.
\end{equation}
A comparison between an asymmetrically-sized and equally-sized MT-SOP is presented in Fig.~\ref{f:cases_1}. The equal-sized converter (Fig.~\ref{f:cases_1a}) has FIC of 1/3~pu for all feeders, whilst an asymmetric MT-SOP sized as \xxx{\alpha}$=(0.35, 0.2, 0.45)$~pu increases the FIC of Feeder \ff{3} to 0.45~pu (Fig.~\ref{f:cases_1b}). This is at the expense of Feeder \ff{2}, which then only has an FIC of 0.2~pu.

\subsubsection{Capability Chart of a Hybrid MT-SOP}
\begin{figure}\centering
\includegraphics[width=0.44\textwidth]{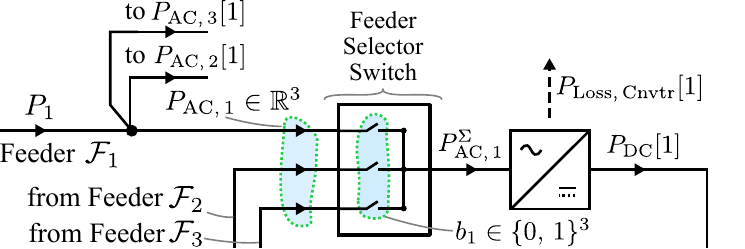}
\caption{Single line diagram of one leg of a Hybrid MT-SOP. The vector $b_{1}$ describes the state of the Feeder Selector Switch. Other variables are used in the operational optimization model, and are fully described in Section~\ref{ss:mop_operational}.}\label{f:pscc_variables}
\end{figure}

Fig.~\ref{f:pscc_variables} shows a single line diagram of one leg of a Hybrid MT-SOP. The state of the Feeder Selector Switch for this leg is denoted by a Selector state vector $b_{i}$, whose $j$th element denotes if the $i$th converter is connected to the $j$th feeder. In other words, the Selector vector $b_{i}$ can be described by
\begin{align}
b_{i} &\in \{0,1\}^{3}\quad i \in \{1,2,3\}\,, \label{e:relay_1} \\
\sum b_{i}&=1\quad i \in \{1,2,3\}\,.\label{e:relay_2}
\end{align}
For example, if converter 1 is connected to feeder $\mathcal{F}_{2}$, then $b_{1}=[0,1,0]^{\intercal}$. As a result, for the $j$th feeder, the three power injection constraints \eqref{e:individuals} can be replaced with the new constraint
\begin{equation}\label{e:new_capability_constraint}
|P_{j}| \leq \sum_{i=1}^{3} b_{i}[j]\alpha_{i} \quad \forall \: j \in \{1,2,3\}\,,
\end{equation}
with the total capability chart found by considering the union of all of the capabilities across all possible Feeder Selector Switch states. 

The capability for the Hybrid MT-SOP that was considered for the previous asymmetric case is shown in Fig.~\ref{f:cases_1c}. Because any converter can be connected to any feeder, there are two lines of symmetry, about $P_{1}=P_{2}$ and $P_{1}=-P_{2}$. The increase in FIC is particularly marked for Feeder \ff{2}, which has increased from 0.2~pu to 0.45~pu. Note that the range of feasible power flows is much wider than the asymmetric case. Additionally, the capability chart of the equally-sized MT-SOP (Fig.~\ref{f:cases_1a}) is a proper subset of the Hybrid MT-SOP capability chart (Fig.~\ref{f:cases_1c}).

\subsubsection{FIC for a given Hybrid MT-SOP state}
For the Hybrid MT-SOP with given Selector state vectors $b_{i}$, the \textit{effective} converter capability for the $j$th feeder $\hat{\alpha}_{j}$ is given by the sum of connected converters to that feeder
\begin{equation}\label{e:alpha_hat}
\hat{\alpha}_{j} = \sum_{i=1}^{3} b_{i}[j]\alpha_{i}\,.
\end{equation}
The FIC for Hybrid MT-SOP designs in a fixed configuration can therefore be calculated by replacing $ \alpha_{j} $ with $ \hat{\alpha}_{j} $ in \eqref{e:fic_calc}.

\subsection{Hybrid MT-SOP Designs}\label{ss:cases}
\begin{figure}\centering
\includegraphics[width=0.49\textwidth]{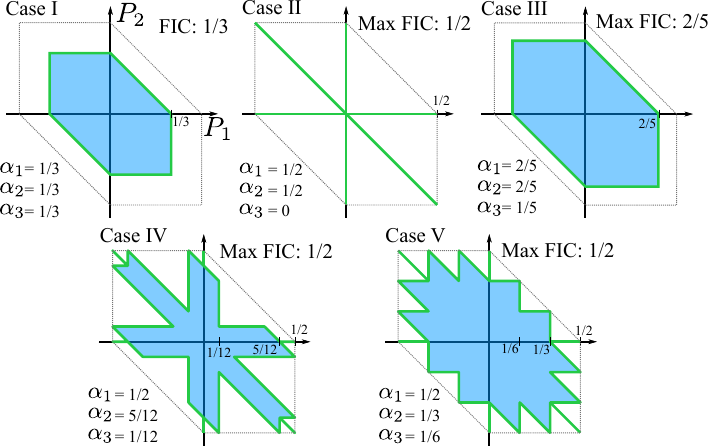}
\caption{Capability charts for five Hybrid MT-SOP designs. By changing the relative sizes of the converters \xxx{\alpha}, the flexibility of the device can be increased compared to a device with equally-size converters (Case~I). Each capability chart is bounded by  $|P_{j}|<1/2 \: \forall \: j \in \{1,2,3\}$, represented by the thin dashed line.}\label{f:cases}
\end{figure}

To further demonstrate how Hybrid MT-SOP designs can change the flexibility that can be afforded by the device, we consider five further combinations of \xxx{\alpha} in Fig.~\ref{f:cases}. The following observations can be made.
\begin{itemize}
\item The two-converter Hybrid MT-SOP (Case~II) has an increased maximum FIC compared to the balanced case (Case~I), but only two feeders can have non-zero FIC at any one time.
\item The capability chart of Case~III is a linear enlargement of the capability chart of Case~I.
\item Case II is a proper subset of each of Cases IV and V.
\item The maximum FIC of Case~III is smaller than that of Cases IV-V, but the capability chart is not a proper subset of either design (e.g., consider around (2/5,0). Additionally, the capability chart is convex.
\end{itemize}
Which Hybrid MT-SOP design is optimal will depend on the situation in which it will be used. In some cases, being able to transfer the full 0.5~pu between feeders may be necessary, whilst for others a convex capability chart may be preferred. 

\section{Operational System Model}\label{s:mop_operation}

The goal of the Hybrid MT-SOP is to increase the utilization of the power electronics, for increasing both the potential power transfers as well as ongoing system-wide operation benefits. In this section, we introduce a conic mixed-integer system optimization model to achieve the latter.

Specifically, in this work, the goal of the optimization is to minimise the cost of energy (i.e., total system losses) by controlling the state of the Feeder Selector Switches and subsequently the power transferred between feeders. These total system losses $P_{\mathrm{Loss,\,Tot}}$ are given by
\begin{equation}\label{e:obj_val}
P_{\mathrm{Loss,\,Tot}} = P_{\mathrm{Loss,\,Ntwk}} + \sum_{i=1}^{3 }P_{\mathrm{Loss,\,Cnvtr}}[i]\,,
\end{equation}
where $P_{\mathrm{Loss,\,Ntwk}},\,P_{\mathrm{Loss,\,Cnvtr}}$ are the network and converter losses, respectively.

\subsection{Hybrid MT-SOP Model and Network Loss Model}\label{ss:mop_operational}

Each leg of the Hybrid MT-SOP is connected to each feeder, with the input of each switch of the Feeder Selector Switch represented by the vector $P_{\mathrm{AC},\,j} \in \mathbb{R}^{3}$ (as shown previously in Fig.~\ref{f:pscc_variables}). The power flow into the converter of the $j$th leg, $P_{\mathrm{AC},\,j}^{\Sigma} \in \mathbb{R}$, is the sum of these AC input powers,
\begin{equation}\label{e:sigma}
P_{\mathrm{AC},\,j}^{\Sigma} = \sum _{i=1}^{3} P_{\mathrm{AC},\,j}[i]\,,
\end{equation}
with the switching operation enforced with the constraints
\begin{equation}\label{e:enforcing_sigma}
|P_{\mathrm{AC},\,j}[i]| \leq b_{j}[i]\alpha_{j} P^{+} \quad \forall \: i \in \{1,2,3\}\,.
\end{equation}
In these expressions, the Selector state vector $ b_{j} $ is as described in \eqref{e:relay_1}, \eqref{e:relay_2}.

The nodal power flow balance at the DC node is denoted
\begin{equation}\label{e:nodal_dc}
\sum _{j=1}^{3} P_{\mathrm{DC}}[j] = 0\,,
\end{equation}
and at the $i$th Feeder point of common coupling,
\begin{equation}\label{e:nodal_feeder}
P_{\mathrm{AC},\,1}[i]+P_{\mathrm{AC},\,2}[i]+P_{\mathrm{AC},\,3}[i] = P_{i} \quad \forall\, i \in \{1,\,2,\,3\} \,.
\end{equation}

The power balance across the $j$th AC/DC converter, considering losses $P_{\mathrm{Loss,\,Cnvtr}}$, is denoted
\begin{equation}\label{e:nodal_acdc}
P_{\mathrm{DC}}[j] + P_{\mathrm{Loss,\,Cnvtr}}[j] = P_{\mathrm{AC},\,j}^{\Sigma}\,.
\end{equation}
For the purposes of this work, converter losses are assumed to be linear with real power transfer (as considered in previous works on SOPs, e.g., \cite{lou2020new,li2017optimal}). For the $i$th converter, we calculate converter losses $P_{\mathrm{Loss,\,Cnvtr}}[i]$ as
\begin{equation}\label{e:conv_loss}
P_{\mathrm{Loss,\,Cnvtr}}[i] = \kappa|P_{\mathrm{DC}}[i]|\,,
\end{equation}
where $\kappa$ is the converter loss coefficient.

We use of a quadratic model for distribution network losses as a function of Hybrid MT-SOP power injections $P_{\mathrm{inj}}$,
\begin{equation}\label{e:ntwk_loss}
P_{\mathrm{Loss,\,Ntwk}} = P_{\mathrm{inj}}^{\intercal}Q_{\mathrm{Ntwk}}P_{\mathrm{inj}} + q_{\mathrm{Ntwk}}^{\intercal}P_{\mathrm{inj}} + c_{q},\,
\end{equation}
with $Q_{\mathrm{Ntwk}},\,q_{\mathrm{Ntwk}},\,c_{Ntwk}$ the quadratic model coefficients, and $P_{\mathrm{inj}}=[P_{1},\,P_{2},\,P_{3}]^{\intercal}$ the vector of power transfers. To derive this, we use the approach previously used in \cite{deakin2020control}. This first calculates the Jacobian of the complex voltages in power injections (using the `First Order Taylor' method of \cite{bernstein2017linear}). Subsequently, the primitive admittances of all individual branch elements are used to derive the quadratic loss model \eqref{e:ntwk_loss} by summing the losses of each element \cite{deakin2020control}.

\subsection{Formulation as a Conic Mixed-Integer Program}\label{ss:formulation}

To formulate the MT-SOP and Network Loss model as a conic mixed-integer program, we must combine the converter power balance \eqref{e:nodal_acdc} and converter loss model \eqref{e:conv_loss} into a suitable form, as well as converting the (convex) quadratic loss constraint \eqref{e:ntwk_loss} into a conic form.

To achieve the latter, the objective function \eqref{e:obj_val} can be rewritten
\begin{equation}\label{e:obj_reform}
P_{\mathrm{Loss,\,Tot}} = \tau _{\mathrm{Ntwk}} + q_{\mathrm{Ntwk}}^{\intercal}P_{\mathrm{inj}} + c_{q} + \sum_{i=1}^{3 }P_{\mathrm{Loss,\,Cnvtr}}[i]\,.
\end{equation}
The dummy variable $\tau _{\mathrm{Ntwk}}$ is introduced alongside the relaxed rotated second order cone constraint (see, e.g., \cite{mosek2021conic})
\begin{equation}\label{e:conic_quad}
\|H_{\mathrm{Ntwk}}P_{\mathrm{inj}} \|_{2}^{2}\leq \tau_{\mathrm{Ntwk}}\,,
\end{equation}
where $ H_{\mathrm{Ntwk}} $ is the Cholesky decomposition of $ Q_{\mathrm{Ntwk}} $.

The power balance of the $i$th AC/DC converter \eqref{e:nodal_acdc} can be rewritten (using \eqref{e:conv_loss}) as
\begin{equation}
P_{\mathrm{AC},\,j}^{\Sigma} = \max \{(1 - \kappa)P_{\mathrm{DC},\,j},\,(1 + \kappa)P_{\mathrm{DC},\,j} \}\,.
\end{equation}
This constraint can be enforced using the big-$M$ method (see, e.g., \cite{mosek2021cookbook}). Specifically, converter losses can be enforced by introducing further binary variables $ z_{j,\,1},\, z_{j,\,2} $ for the $j$th converter such that
\begin{align}
(1 - \kappa)P_{\mathrm{DC},\,j} &\leq P_{\mathrm{AC},\,j}^{\Sigma} \leq (1 - \kappa)P_{\mathrm{DC},\,j} + M(1-z_{j,\,1})\,, \label{e:bigM1}\\
(1 + \kappa)P_{\mathrm{DC},\,j} &\leq P_{\mathrm{AC},\,j}^{\Sigma} \leq (1 + \kappa)P_{\mathrm{DC},\,j} + M(1-z_{j,\,2})\,, \\
z_{j,\,1} & + z_{j,\,2} = 1\,,\\
z_{j,\,1} & \in \{0,\,1\} \,, \quad z_{j,\,2} \in \{0,\,1 \} \,.\label{e:bigM4}
\end{align}

The operational optimization model of the network-Hybrid MT-SOP system can therefore be summarised by the optimization
\begin{align}
\min \: & \eqref{e:obj_reform}\,, \label{e:opt_of} \\
\mathrm{s.t.} \: \eqref{e:sigma}-\eqref{e:nodal_feeder},\,&\,\eqref{e:ntwk_loss},\,\eqref{e:conic_quad},\,\eqref{e:bigM1}-\eqref{e:bigM4}\,. \label{e:opt_cons}
\end{align}
This optimization is implemented using Mosek Fusion \cite{mosek2021mosek}. OpenDSS \cite{opendss2021} is used as a power flow solver, interfaced through DSS Extensions \cite{meira2021dss}.

\subsection{Evaluating Hybrid MT-SOP Performance}\label{ss:performance}

To evaluate the operational performance of the Hybrid MT-SOP, we consider two metrics. Firstly, the change in the value of the objective function \eqref{e:obj_reform} summarizes the improved performance that the additional flexibility brings. We also consider the utilization $\eta$, evaluated as
\begin{equation}\label{e:util_defn}
\eta = \dfrac{1}{T} \sum _{t=1}^{T} \dfrac{\|P(t)\|_{1}}{P^{+}}\,,
\end{equation}
where $P^{+}$ is the total device rating and $T$ is the total number of time periods $t$ over which the utilization is calculated. Higher utilization rates indicate a power converter that is transferring more power (on average) between feeders over a given period.

\section{Case Studies}\label{s:casestudy}

To demonstrate and compare the benefits afforded by the flexible interconnection that Hybrid MT-SOPs provide, the five designs of Section~\ref{ss:cases} are used in this section as the basis for a case study in a future distribution system. The case study is introduced in Section~\ref{ss:setup}, with the solutions analysed in detail in Section~\ref{ss:results}.

\subsection{Case Study Setup}\label{ss:setup}

\begin{figure}\centering
\subfloat[33 Bus Network]{\includegraphics[width=0.32\textwidth]{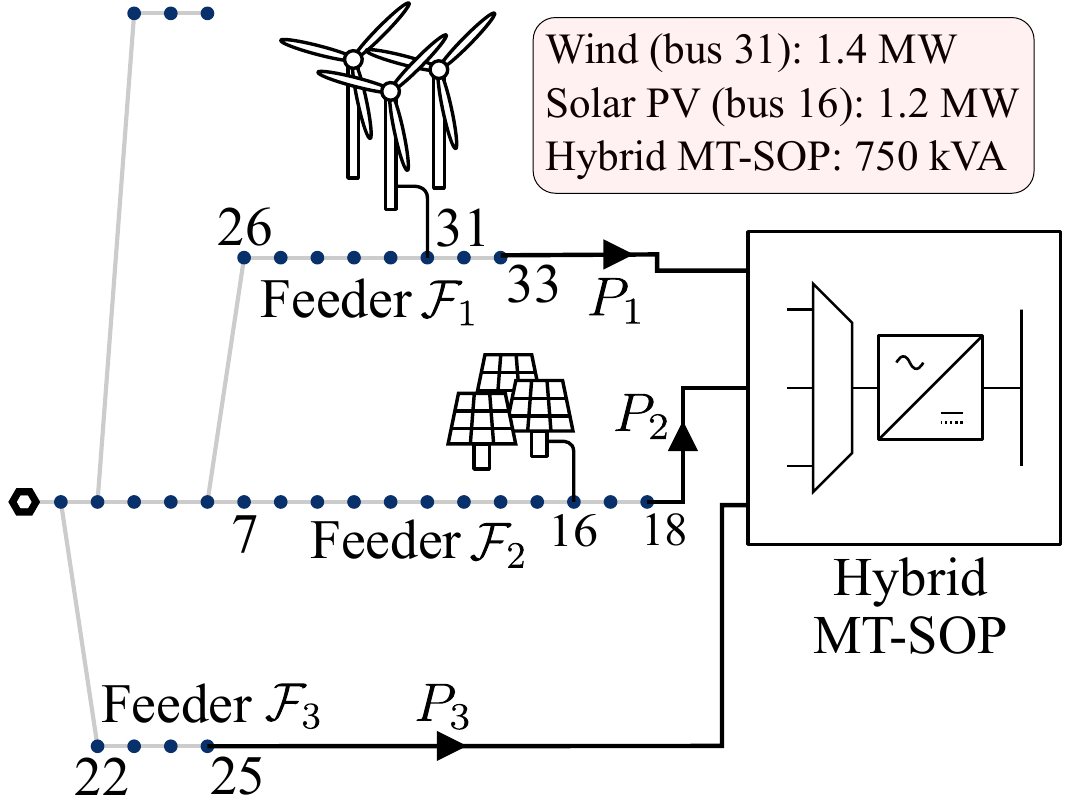}\label{f:33bus}}
~
\subfloat[Profiles]{\includegraphics[width=0.165\textwidth]{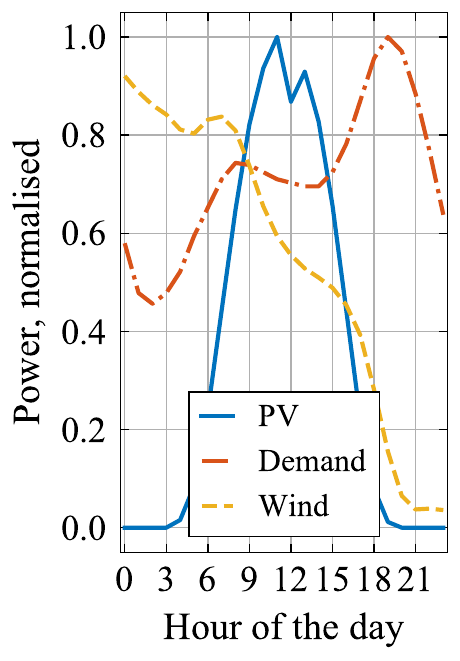}\label{f:pltDemandProfiles}}
\caption{(a) The case study is based on the 33 Bus network of Baran and Wu \cite{baran1989network} with a 1.4 MW wind farm, 1.2 solar PV park, and 750~kVA Hybrid MT-SOP. (b) The demand, solar and wind profiles used in the case studies.}
\label{f:case_study_basics}
\end{figure}

The case study is intended to consider a future network with a high penetration of LCTs. We therefore modify the Baran \& Wu 33-Bus network to include LCTs (Fig.~\ref{f:case_study_basics}), with a 1.4 MW wind generator and 1.2 MW solar PV generator connected at buses 31 and 16 respectively (by comparison, the total load of the network is 3.7 MW). A Hybrid MT-SOP is connected between buses 33, 18, and 25 (as in \cite{ji2017enhanced}), with a feeder-feeder power transfer of 98\%, such that the loss coefficient $\kappa = 1\%$ \cite{cao2016benefits}.

A representative wind generator capacity factor is taken from \cite{bloomfield2020merra} (for 24th June 2013). A solar PV profile is based on Typical Meteorological Year data from \cite{dobos2014pvwatts}, using London Heathrow as a geographic location. The demand profile is based on the residential profile of \cite{short2014electric}[Ch. 2]. Without the Hybrid MT-SOP, these profiles and generator sizes result in reasonable network performance, with the greatest voltage rise of 2.5\% and a 7.5\% voltage drop at peak load.

\subsection{Results}\label{ss:results}

In this section, we first consider the performance of the device in terms of the objective function and solution quality, to demonstrate both the benefits of the proposed Hybrid MT-SOP as well as the suitability of the conic mixed-integer formulation. We then explore the operation of the Hybrid MT-SOP in detail, to highlight how the device has improved the performance of the system.

\subsubsection{Optimal Solutions and Solution Quality}

\begin{figure}\centering
\subfloat[System losses (from \eqref{e:obj_val}) ]{\includegraphics[width=0.195\textwidth]{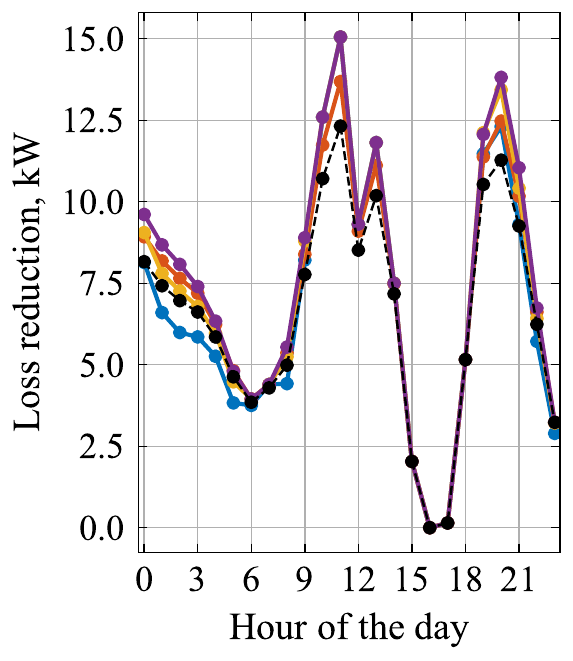}\label{f:pltCaseStudies_lossChange}}
~
\subfloat[Ratio of system losses to Case I]{\includegraphics[width=0.26\textwidth]{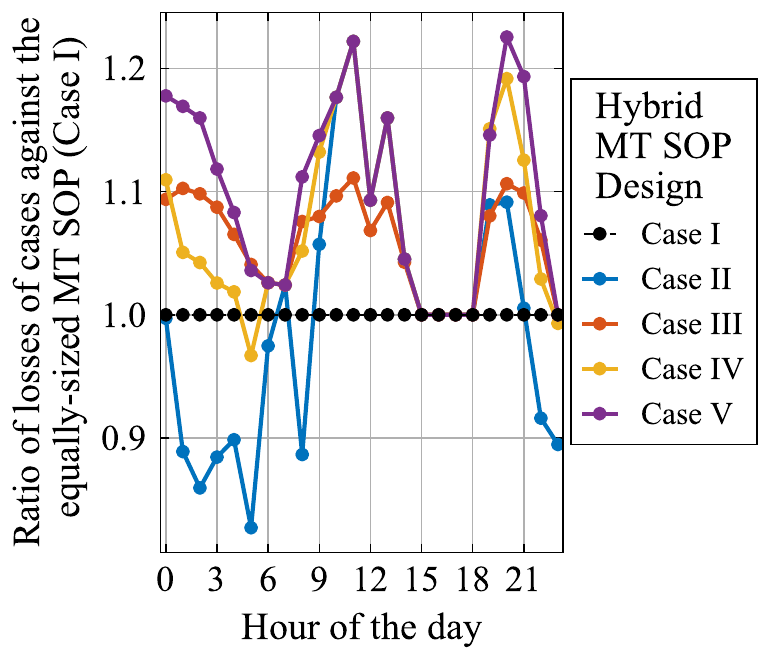}\label{f:pltCaseStudies_lossRatio}}
\caption{System losses (a), and their ratio against the equally-sized MT-SOP (b), for each hour and for each Hybrid MT-SOP design. The Hybrid MT-SOP can have more than 20\% additional loss reduction as compared the equally-sized MT-SOP (Case~I). For comparison, the total losses over the day are 1781 kWh for Case~I.}
\label{f:losses}
\end{figure}

Fig.~\ref{f:losses} plots the reduction in system losses across the day for all five Hybrid MT-SOP designs, including the ratio of losses against the equally-sized MT-SOP (Case~I). It is observed that the hourly loss reduction can be increased by more than 20\% (Fig.~\ref{f:pltCaseStudies_lossRatio}). It is interesting to note that the equally-sized MT-SOP (Case~I) out-performs some Hybrid MT-SOP designs at different times of day (particularly, Case~II in the early morning). Case V clearly has the best performance over the day, as highlighted in Table \ref{t:tblPerformanceSummary}. This table also reports the utilization $\eta$ of the 750~kVA power converters, as calculated according to \eqref{e:util_defn}. Clearly, there is significant scope for improving both the utilization and losses.

\begin{table}
\centering
\caption{Comparing Hybrid MT-SOP (Cases II-V) and equally-sized MT-SOP (Case I) performance, in terms of loss reduction $\Delta P_{\mathrm{Loss,\,Tot}}$ \eqref{e:obj_val} and power converter utilization $\eta$ \eqref{e:util_defn}.}\label{t:tblPerformanceSummary}
\begin{tabular}{lll}
\toprule
Design & \begin{tabular}[x]{@{}l@{}}Loss Reduction\\ $\Delta P_{\mathrm{Loss,\,Tot}}$, kWh/day \end{tabular} & Utilization $\eta$, \% \\
\midrule
Case I & 157.3 & 60.4\% \\
Case II & 161.7 (+2.8\%) & 72.0\% (+19.2\%) \\
Case III & 169.4 (+7.7\%) & 70.8\% (+17.4\%) \\
Case IV & 172.9 (+9.9\%) & 79.1\% (+31.1\%) \\
Case V & 178.2 (+13.3\%) & 82.0\% (+35.9\%) \\
\bottomrule
\end{tabular}

\label{t:tblPerformanceSummary}
\end{table}

The solution quality of the formulation, described in Section~\ref{ss:formulation}, was found to be good. Across all cases and hours, the maximum mixed-integer relative gap was no greater than $10^{-4}$ (with the maximum mixed-integer absolute gap being less than 1~Watt). The rotated second order cone relaxation \eqref{e:conic_quad} was found to be exact to a relative accuracy of $10^{-5}$ across all simulations. The quadratic losses model employed \eqref{e:ntwk_loss} also was found to have good accuracy, with the relative difference between the losses calculated using the non-linear power flow solution and quadratic loss model being less than 1\% across all hours and cases.

\subsubsection{Properties of the Optimal Solution}

\begin{figure*}\centering
\subfloat[Powers, Case~I]{\includegraphics[width=0.195\textwidth]{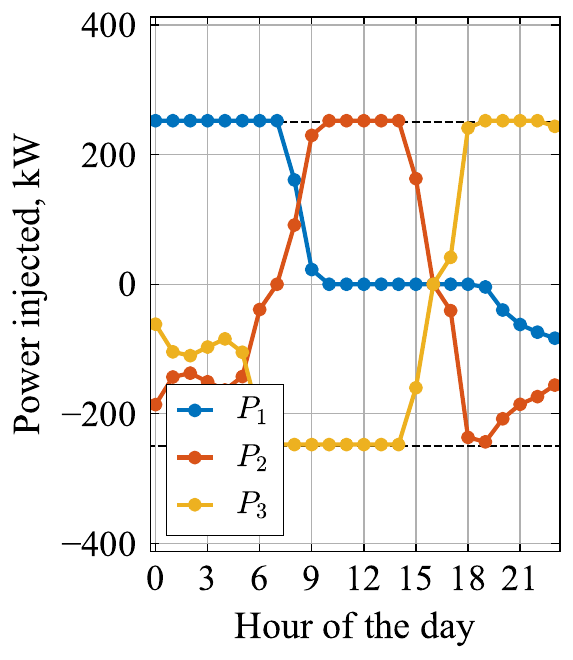}\label{f:powers_I}}
~
\subfloat[Powers, Case~II]{\includegraphics[width=0.195\textwidth]{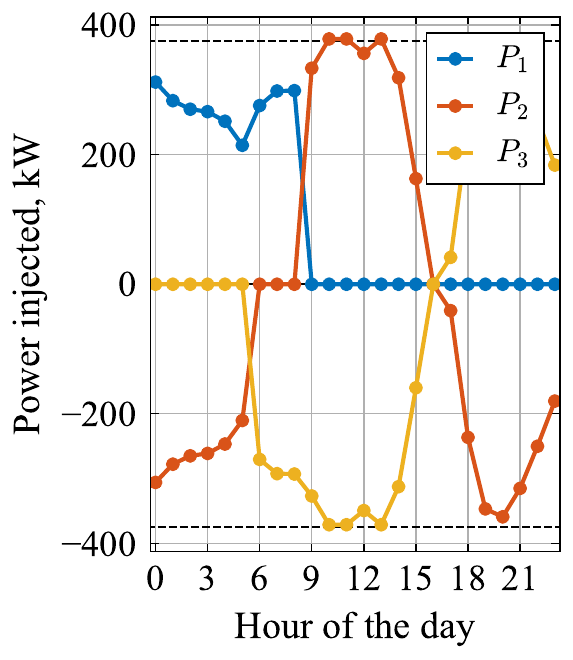}\label{f:powers_II}}
~
\subfloat[Powers, Case~III]{\includegraphics[width=0.195\textwidth]{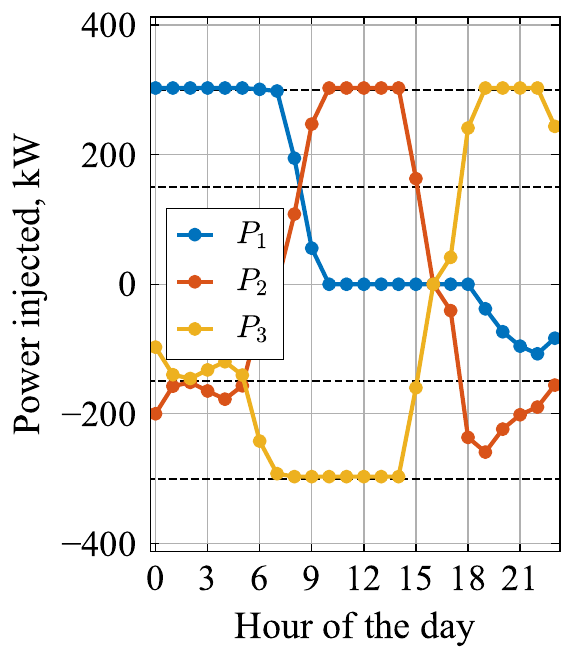}\label{f:powers_III}}
~
\subfloat[Powers, Case~IV]{\includegraphics[width=0.195\textwidth]{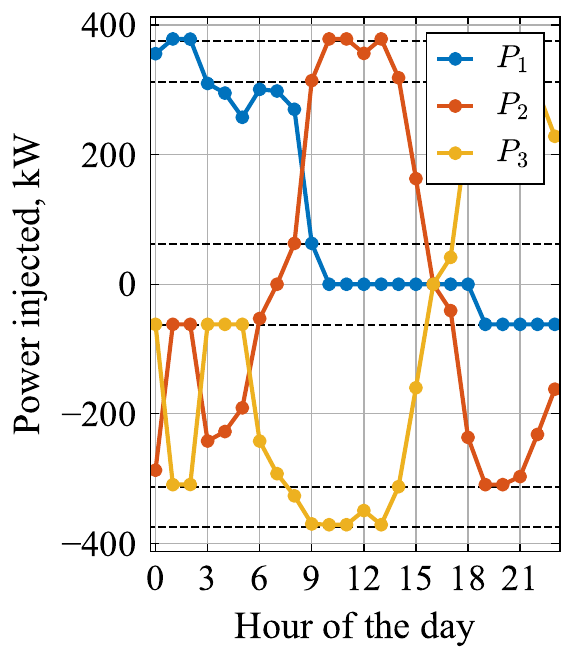}\label{f:powers_IV}}
~
\subfloat[Powers, Case V]{\includegraphics[width=0.19\textwidth]{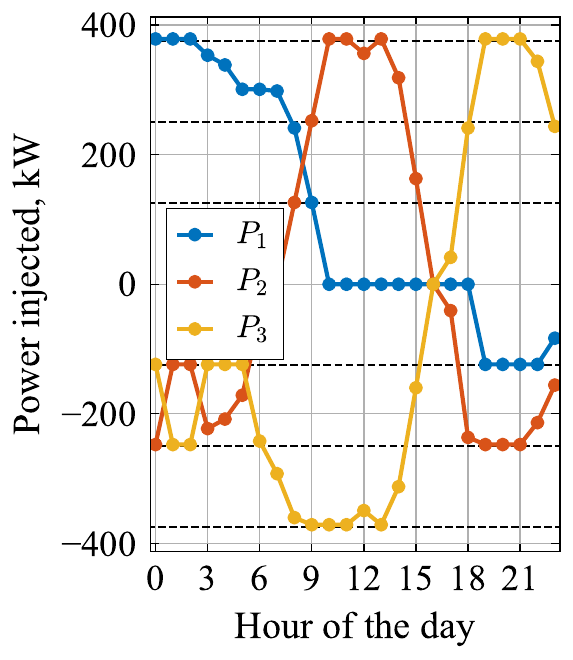}\label{f:powers_V}}
\caption{The optimal powers across the day for the five Hybrid MT-SOP designs (from Fig.~\ref{f:cases}). The general direction of transfers between feeders does not change too much, but the total power that is transferred between feeders does varies significantly.}
\label{f:results}
\end{figure*}

The optimization solutions, in terms of the power injected from each feeder into the Hybrid MT-SOP, are plotted in Fig.~\ref{f:results}. Although the power flows differ, it can be seen that across all cases, there are four distinct phases:
\begin{itemize}
\item Early morning (midnight to 6am). During this period, wind output is high and demand is low. Power is therefore exported from Feeder \ff{1}.
\item Morning through early afternoon (7am to 3pm). As solar PV picks up and wind drops, Feeder \ff{2} exports power (mostly to meet the demand of Feeder \ff{3}).
\item Late afternoon (4-5pm). Wind and solar PV are still meeting their demand locally, and so the Hybrid MT-SOP utilization $\eta$ is very low.
\item Wind and solar generation is very low in the evening (6pm to midnight), and so the peak demands on Feeders \ff{1} and \ff{2} cannot be met locally. The low-impedance Feeder \ff{3} therefore exports power to those feeders.
\end{itemize}
It can be seen that all three feeders import and export at some point in the day across four of the five cases (the exception being Case~II).

Whilst all five cases follow the same diurnal pattern in terms of loss reduction (Fig.~\ref{f:losses}), it can be observed that the optimal solutions change significantly (Fig.~\ref{f:results}). For example, Case~II only has two non-zero powers at a given time, resulting in the lower performance for some hours as compared to the equally-sized Case~I (Fig.~\ref{f:pltCaseStudies_lossRatio}), demonstrating that the increase in the maximum FIC from 250~kVA to 375~kVA is not the only metric by which a Hybrid MT-SOP should be evaluated. Indeed, Case~III has a lower utilization rate $\eta$ than Case~II, but a higher loss reduction through the day (Table \ref{t:tblPerformanceSummary}). Conversely, Case~V has both a higher utilization and loss reduction than Case~IV.

\begin{figure}\centering
\subfloat[FIC, Case~II]{\includegraphics[width=0.24\textwidth]{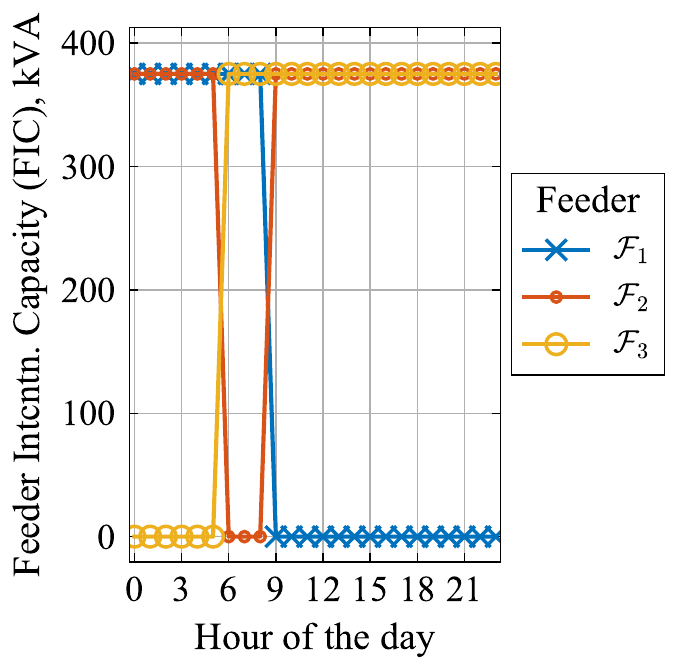}\label{f:capcon_II}}
~
\subfloat[FIC, Case~V]{\includegraphics[width=0.24\textwidth]{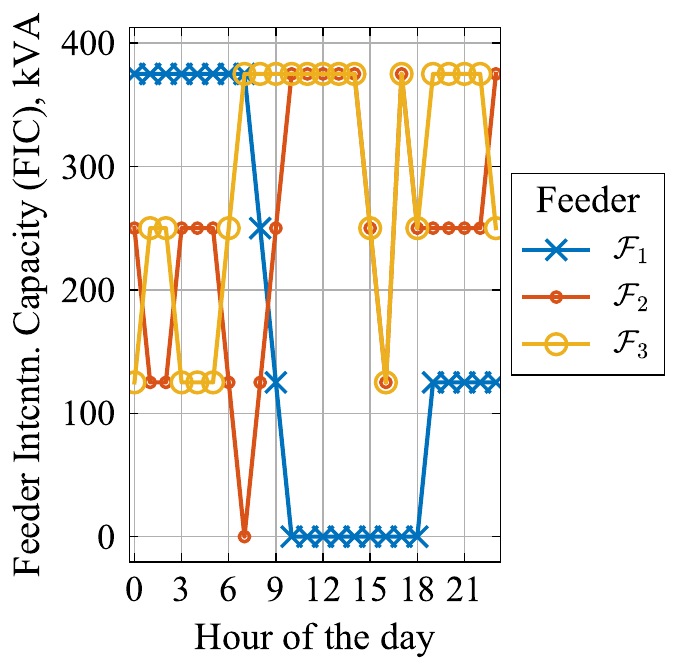}\label{f:capcon_V}}
\caption{The FIC for Case~II (a) and Case V (b), as calculated using \eqref{e:fic_calc} and \eqref{e:alpha_hat}. This demonstrates the effect the optimal configuration of the Hybrid MT-SOP has on the power transfers that can be achieved.}
\label{f:capcon}
\end{figure}

Fig.~\ref{f:capcon} illustrates the impact of the Feeder Selector Switches on device operation by plotting the FIC (calculated from \eqref{e:fic} and \eqref{e:alpha_hat}) for Case~II and Case V. Whilst these both have the same maximum FIC, the FIC across the day is very different for these two designs. The FICs of Case~V change much more frequently through the day as compared to Case~II, and can take four values rather than two. The lifetime of some AC switch designs is sometimes limited to a given number of total switching operations, and so future work could consider the trade-off between reduced converter switching operations and increased system losses.

\section{Conclusions}\label{s:conclusions}

As distribution networks become host to a menagerie of low-carbon technologies, solutions that are optimal from a whole-systems perspective will be required to provide flexibility for increased network capacity and reduced operational costs. This paper introduced the Hybrid Multi-Terminal Soft Open Point to address this issue, using asymmetric converter sizing and Feeder Selector Switches. The locus of feasible power transfers were plotted as device capability charts, demonstrating up to 50\% increase in the feeder interconnection capacity as compared to an equally-sized Multi Terminal SOP. A conic mixed-integer program was developed to minimize the total system losses. Case studies demonstrated the total power converter utilization can be increased by more than one third, with loss reduction increased by up to 13\%.

The focus of this paper has been on demonstrating the Hybrid MT-SOP concept, and so has explored key design and operational issues of the device. Future research could consider the optimal design of a Hybrid MT-SOP for a given network, device design considering any number of feeders and converters, or operational approaches to minimize the number of switching operations.

The relatively high cost of power converters means that high utilization and performance of installed converter capacity is crucial, even when these devices are used in a targeted way to support legacy AC systems. It is concluded that new and novel Hybrid device topologies, such as the Hybrid Multi-Terminal Soft Open Point, could become key technologies to enable efficient and highly utilized distribution systems on the path to Net-Zero.



%
%
%

\bibliography{refs_mops}{}
\bibliographystyle{IEEEtran}

\end{document}